# TENTATIVE APPRAISAL OF COMPATIBILITY OF SMALL-SCALE CMB ANISOTROPY DETECTIONS IN THE CONTEXT OF COBE-DMR-NORMALIZED OPEN AND FLAT Λ CDM COSMOGONIES

Ken Ganga[1], Bharat Ratra[2], and Naoshi Sugiyama[3]

[1]Division of Physics, Mathematics and Astronomy, California Institute of Technology, Pasadena, CA 91125

[2]Center for Theoretical Physics, Massachusetts Institute of Technology, Cambridge, MA 02139

[3]Department of Physics and Research Center for the Early Universe, University of Tokyo, Tokyo 113

## ABSTRACT

Goodness-of-fit statistics are used to quantitatively establish the compatibility of CMB anisotropy predictions in a wide range of DMR-normalized, open and spatially-flat Λ, CDM cosmogonies with the set of all presently available small-scale CMB anisotropy detection data. Conclusions regarding model viability depend sensitively on the prescription used to account for the $1\sigma$ uncertainty in the assumed value of the DMR normalization, except for low-density, $\Omega_0 \sim 0.3 - 0.4$, open models which are compatible with the data for all prescriptions used. While large baryon-density ($\Omega_B \gtrsim 0.0175 h^{-2}$), old ($t_0 \gtrsim 15 - 16$ Gyr), low-density ($\Omega_0 \sim 0.2 - 0.4$), flat-Λ models might be incompatible, no model is incompatible with the data for all prescriptions. In fact, some open models seem to fit the data better than should be expected, and this might be an indication that some error bars are mildly overconservative.

*Subject headings:* cosmic microwave background — cosmology: observations — large-scale structure of the universe

## 1. INTRODUCTION

Recent and near-future measurements of CMB anisotropy, on a variety of angular scales, when used in conjunction with the predicted anisotropy in cosmogonical models, are in the process of transforming the CMB anisotropy field from that in which one tries to draw qualitative conclusions about the viability of broad-brush cosmological models, to that in which it will soon be possible to set quantitative constraints on parameters of some specific models and rule out other models. Until now all quantitative comparisons between model predictions and the data have made use of one of two simplifications: (1) data from one (or a few) experiments has been compared to predictions for one (or a class of) model(s); or, (2) data from a larger number of experiments has been compared to predictions for a single model. While clearly a necessary first step, this approach has led to a number of vague claims about the (in)compatibility of some model(s) with some subset of the data, which, while perhaps correct, need to be put on firmer ground.

This work is a first attempt to compare all presently available CMB anisotropy detection data to predictions in a wide variety of observationally motivated cosmogonies, with the ultimate goal of deciding, in a quantitative manner, whether (or if) any of these models are compatible with the wealth of CMB anisotropy data. Such a quantitative approach, using all available data, is essential if one wishes to draw robust conclusions about model viability. It will become more effective as the analyses are understood better, and as the data improves. The only other alternative is to wait a decade or so for a new CMB anisotropy satellite to address this issue.

To qualitatively assess compatibility, Ratra et al. (1995, hereafter RBGS) and Ratra & Sugiyama (1995, hereafter RS) compared anisotropy predictions in 2 year DMR-normalized, gaussian, adiabatic, open and spatially-flat $\Lambda$, CDM models (with the values of $\Omega_0$, $h$, and $\Omega_B$ chosen to satisfy non-CMB observational constraints, except in the fiducial CDM case) to small-scale anisotropy data[4]. Here we use these predictions, in combination with a variety of goodness-of-fit statistics[5], to quantitatively assess the compatibility of CMB anisotropy detections. In contrast to RBGS and RS, we explore more options for accounting for the $1\sigma$ uncertainty in the DMR normalization, but in this preliminary analysis we ignore small-scale CMB anisotropy upper limits as well as the small correlations between data points from experiments with multiple windows (see §3).

---

[4] It is important to bear in mind that a variety of statistical techniques and prescriptions (as well as different assumed CMB anisotropy spectra) have been used to determine the observational results (RBGS), that the usual prescription for accounting for calibration uncertainty, by adding it in quadrature, is not quite correct (RBGS), and that non-CMB contamination and/or subtraction might be an issue in some cases.

In a related analysis, Scott, Silk, & White (1995, hereafter SSW) used a Lorentzian approximation for the shape of the fiducial CDM model CMB anisotropy spectrum and concluded that it provided an adequate description of the anisotropy data (they took the data error bars to be symmetric). Here we use significantly more observational data, as well as revised estimates of some of the older data, and also use numerically computed CMB anisotropy spectra for a wider variety of models motivated by non-CMB observations. Consistent with the conclusion of SSW, for all prescriptions we have used to account for the allowed $1\sigma$ range of the DMR normalization, low-density open models with $\Omega_0 \sim 0.3$ – 0.4 are compatible with the data, so a CMB anisotropy spectrum that mildly rises to multipole moments $l \sim 200$ is compatible with the data[6].

## 2. SUMMARY OF COMPUTATION

We consider 32 smaller-scale CMB anisotropy detections (almost entirely sensitive only to $l \lesssim 200$): FIRS (Ganga et al. 1994); Tenerife (Hancock et al. 1995); SK93 and individual-chop SK94 Ka and Q (Netterfield et al. 1996); SP94 Ka and Q (Gundersen et al. 1995); Python-G, -L, and -S (e.g., Platt et al. 1995); ARGO (de Bernardis et al. 1994); MAX3, individual-channel MAX4, and MAX5 (e.g., Tanaka et al. 1995); MSAM92 and MSAM94 (Cheng et al. 1995); and WDH1 (Griffin et al. 1995). For the $i^{\rm th}$ detection, the observed bandtemperature $\delta T^i_{(e)}$ and the $1\sigma$ upper and lower limits, $\delta T^i_{(e)u}$ and $\delta T^i_{(e)l}$, and the DMR-normalized, 26 open and flat-$\Lambda$, model predictions for the bandtemperature $1\sigma$ (gaussian) upper and lower limits, $\delta T^i_{(m)u}$ and $\delta T^i_{(m)l}$ (range accounts for statistical and systematic uncertainty in the DMR normalization, Stompor, Górski, & Banday 1995), are given in RBGS (open models, DMR-galactic-frame normalization) and RS (flat-$\Lambda$ models, DMR-ecliptic-frame normalization). Details may be found in these papers, and the model parameter values are listed in Table 1.

---

[5] We use various such statistics since most observational error bars are asymmetric (i.e., nongaussian).

[6] As noted by SSW, such a rising CMB anisotropy spectrum is consistent with that expected from radiation-pressure-induced oscillations at early times in the adiabatic structure formation picture. It is also consistent with that expected in versions of the isocurvature scenario. And there almost certainly are other models that are consistent with the data.

To assess the effect of varying the DMR normalization, we consider 3 sets of model predictions $\delta T^i_{(m)c}$, normalized to the lower $1\sigma$ ($= \delta T^i_{(m)l}$), central ($= [\delta T^i_{(m)u} + \delta T^i_{(m)l}]/2$), and upper $1\sigma$ ($= \delta T^i_{(m)u}$) values of the DMR normalization. For each of these predictions we also either account for or ignore the model (DMR) "errors" $\sigma^i_{(m)} = [\delta T^i_{(m)u} - \delta T^i_{(m)l}]/2$, yielding 6 sets of predictions. Five prescriptions ($I$) are used to construct the corresponding observational numbers: (1) $\delta T^i_{(e)c} = \delta T^i_{(e)}$ and $\sigma^i_{(e)} = \delta T^i_{(e)u} - \delta T^i_{(e)}$; (2) $\delta T^i_{(e)c} = \delta T^i_{(e)}$ and $\sigma^i_{(e)} = [\delta T^i_{(e)u} - \delta T^i_{(e)l}]/2$; (3) $\delta T^i_{(e)c} = [\delta T^i_{(e)u} + \delta T^i_{(e)l}]/2$ and $\sigma^i_{(e)} = [\delta T^i_{(e)u} - \delta T^i_{(e)l}]/2$; (4) $\delta T^i_{(e)c} = \delta T^i_{(e)}$ and $\sigma^i_{(e)}$ either $= [\delta T^i_{(e)u} - \delta T^i_{(e)c}]$ if $\delta T^i_{(m)c} > \delta T^i_{(e)c}$ or $= [\delta T^i_{(e)c} - \delta T^i_{(e)l}]$ if $\delta T^i_{(m)c} < \delta T^i_{(e)c}$; and, (5) $\delta T^i_{(e)c} = \delta T^i_{(e)}$ and $\sigma^i_{(e)} = \delta T^i_{(e)} - \delta T^i_{(e)l}$. For each model, DMR-normalization value, model "error" prescription, detection $i$, and observational data prescription $I$, $D^i_{(I)} = [\delta T^i_{(e)c} - \delta T^i_{(m)c}]/\sigma^i$ [where $\sigma^i$ is either $= |\sigma^i_{(e)}|$ or $= ([\sigma^i_{(e)}]^2 + [\sigma^i_{(m)}]^2)^{0.5}$] is used as a measure of the deviation of the prediction from the observation. The $D^i_{(I)}$ are used to compute the reduced goodness-of-fit statistic $\chi^2_{(I)} = \sum_{i=1}^{32}[D^i_{(I)}]^2/32$ (usual reduced $\chi^2$ for points drawn from a gaussian distribution). These are shown in Figs. 1 – 6.

## 3. DISCUSSION

Some of the 32 detections we use here are not completely independent. As a result, there are slightly less than the 32 degrees of freedom we have assumed here, which, by itself, causes our reduced $\chi^2_{(I)}$ values to be slightly smaller than they should be. (A more accurate computation will require the appropriate correlation matrices.) One might hope to roughly compensate for this by focussing on the $\chi^2_{(I)}$ computed using $D^i_{(I)}$ with $|\sigma^i_{(e)}|$ in the denominator [instead of $([\sigma^i_{(e)}]^2 + [\sigma^i_{(m)}]^2)^{0.5}$ which leads to an overestimate of the uncertainty due to cosmic variance (which has already been accounted for in the small-scale data error bars, and is an issue especially for data points from larger-angle experiments), and that due to systematic shifts in DMR normalization (which has already been accounted for by our use of 3 different DMR-normalization values)], but this would ignore the DMR noise uncertainty.

The skewness and kurtosis of the 780 $D_{(I)}^i$ distributions, for each model, DMR-normalization value, observational data prescription, and model "error" prescription, is consistent with the range set by the variances of the skewness and kurtosis for 32 degrees of freedom drawn from a gaussian distribution. This means that less-compatible models (large $\chi_{(I)}^2$ in Figures) are less-compatible because of many somewhat deviant predictions, and not because of just a few extremely deviant predictions.

Focussing on the nominal-DMR-normalized $\chi_{(I)}^2$ (Figs. 1 & 2), independent of the model "error" prescription, the only low-density ($\Omega_0 \sim 0.2 - 0.4$) flat-$\Lambda$ models compatible with the data (i.e., with $\chi_{(I)}^2 < 1.46$, which is $2\sigma$ [4.55% probability of $\chi_{(I)}^2$ being this large or larger] for 32 degrees of freedom drawn from a gaussian distribution) are the younger ($t_0 \sim 13$ Gyr), lower baryon-density ($\Omega_B \lesssim 0.0075 h^{-2}$) ones[7], while all open models are compatible with the data (in agreement with the qualitative conclusions of RBGS and RS). For the upper $1\sigma$ value of the DMR normalization (Figs. 3 & 4), independent of the model "error" prescription, all flat-$\Lambda$ models are incompatible, while low-density open models are compatible. At the lower $1\sigma$ value of the DMR normalization (Figs. 5 & 6), the $\Omega_0 = 0.1$ open model is incompatible, most other models are compatible, and, in this case, conclusions regarding the viability of low-density ($\Omega_0 \sim 0.2 - 0.4$), old ($t_0 \sim 15 - 16$ Gyr), high baryon-density ($\Omega_B \sim 0.0175 h^{-2}$) flat-$\Lambda$ models depend sensitively on the model "error" prescription[8].

Independent of the DMR-normalization value and model "error" prescription, low-density, $\Omega_0 \sim 0.3 - 0.4$, open models (models 4 – 9) are compatible with the data. This is our only robust conclusion about model viability. However, this is based on the assumption that there are no gross, unaccounted for, systematic uncertainties, and it only means that, in this case, the quoted error bars are not unreasonably small. (We emphasize that even if there are gross, unaccounted for, systematic uncertainties, the CMB anisotropy detections could still be compatible, but with different models.)

---

[7] It might be significant that for flat-$\Lambda$ models the CMB anisotropy data seems to favour a larger $h$ and a smaller $\Omega_B$, while some large-scale structure observations seem to favour a smaller $h$ and a larger $\Omega_B$ (e.g., Stompor et al. 1995; SSW).

[8] In our analysis here we have ignored upper limits. For the models we consider, the only seriously constraining upper limit is that of WDI (Tucker et al. 1993). This is mostly a serious constraint for the flat-$\Lambda$ models (RS; RBGS), and is probably incompatible with these particular flat-$\Lambda$ models (RS).

It might be significant that a fairly large number of $\chi^2_{(I)}$ are less than unity. While correlations between some data points certainly contribute to this, our understanding of the magnitude of the correlations leads us to suspect that mildly overconservative error bars on some of the data points might also be an issue. If this turns out to be more than just idle speculation, and if it can then be resolved, then, in combination with near-future improved small-scale CMB anisotropy data, and the tighter normalization error bars expected from the 4 year DMR data, our quantitative approach, based on using all available data, should allow for more robust conclusions about model viability.

We are indebted to T. Banday, L. Page and J. Peebles, and also acknowledge useful discussions with E. Bertschinger, K. Górski, G. Griffin, J. Gundersen, B. Netterfield, U. Seljak, and S. Tanaka.

TABLE 1

Numerical Values for Model Parameters

| # | $\Omega_0$ | $h$ | $\Omega_B h^2$ |
|---|---|---|---|
| (O)0 | 0.1 | 0.75 | 0.0125 |
| (O)1 | 0.2 | 0.65 | 0.0175 |
| (O)2 | 0.2 | 0.70 | 0.0125 |
| (O)3 | 0.2 | 0.75 | 0.0075 |
| (O)4 | 0.3 | 0.60 | 0.0175 |
| (O)5 | 0.3 | 0.65 | 0.0125 |
| (O)6 | 0.3 | 0.70 | 0.0075 |
| (O)7 | 0.4 | 0.60 | 0.0175 |
| (O)8 | 0.4 | 0.65 | 0.0125 |
| (O)9 | 0.4 | 0.70 | 0.0075 |
| (O)10 | 0.5 | 0.55 | 0.0175 |
| (O)11 | 0.5 | 0.60 | 0.0125 |
| (O)12 | 0.5 | 0.65 | 0.0075 |
| (O)13 | 1.0 | 0.50 | 0.0125 |
| ($\Lambda$)14 | 0.1 | 0.90 | 0.0125 |
| ($\Lambda$)15 | 0.2 | 0.70 | 0.0175 |
| ($\Lambda$)16 | 0.2 | 0.75 | 0.0125 |
| ($\Lambda$)17 | 0.2 | 0.80 | 0.0075 |
| ($\Lambda$)18 | 0.3 | 0.60 | 0.0175 |
| ($\Lambda$)19 | 0.3 | 0.65 | 0.0125 |
| ($\Lambda$)20 | 0.3 | 0.70 | 0.0075 |
| ($\Lambda$)21 | 0.4 | 0.55 | 0.0175 |
| ($\Lambda$)22 | 0.4 | 0.60 | 0.0125 |
| ($\Lambda$)23 | 0.4 | 0.65 | 0.0075 |
| ($\Lambda$)24 | 0.5 | 0.60 | 0.0125 |
| ($\Lambda$)25 | 1.0 | 0.50 | 0.0125 |

FIGURE CAPTIONS

Fig. 1.– Five reduced (for 32 degrees of freedom) $\chi^2_{(I)}$, computed for the nominal value of the DMR normalization, and accounting for model "errors", as a function of model number. The vertical short-dashed line divides the open models from the flat-$\Lambda$ models. See Table 1 for model numbers and parameter values. The horizontal short-dashed lines are, in ascending order, at $1\sigma$ (31.7% probability of $\chi^2_{(I)}$ being this large or larger), $2\sigma$ (4.55% probability), $3\sigma$ ($2.70 \times 10^{-3}$ probability), $4\sigma$ ($6.33 \times 10^{-5}$ probability), and $5\sigma$ ($5.73 \times 10^{-7}$ probability) for 32 degrees of freedom drawn from a gaussian distribution, and the horizontal long-dashed line is at the 95% probability level.

Fig. 2.– As for Fig. 1, but computed for the nominal value of the DMR normalization, and now ignoring model "errors". Note that the vertical axis scale is different.

Fig. 3.– As for Fig. 1, but now computed for the upper $1\sigma$ value of the DMR normalization, and accounting for model "errors". Note that the vertical axis scale is different.

Fig. 4.– As for Fig. 1, but now computed for the upper $1\sigma$ value of the DMR normalization, and ignoring model "errors". Note that the vertical axis scale is different.

Fig. 5.– As for Fig. 1, but now computed for the lower $1\sigma$ value of the DMR normalization, and accounting for model "errors".

Fig. 6.– As for Fig. 1, but now computed for the lower $1\sigma$ value of the DMR normalization, and ignoring model "errors".

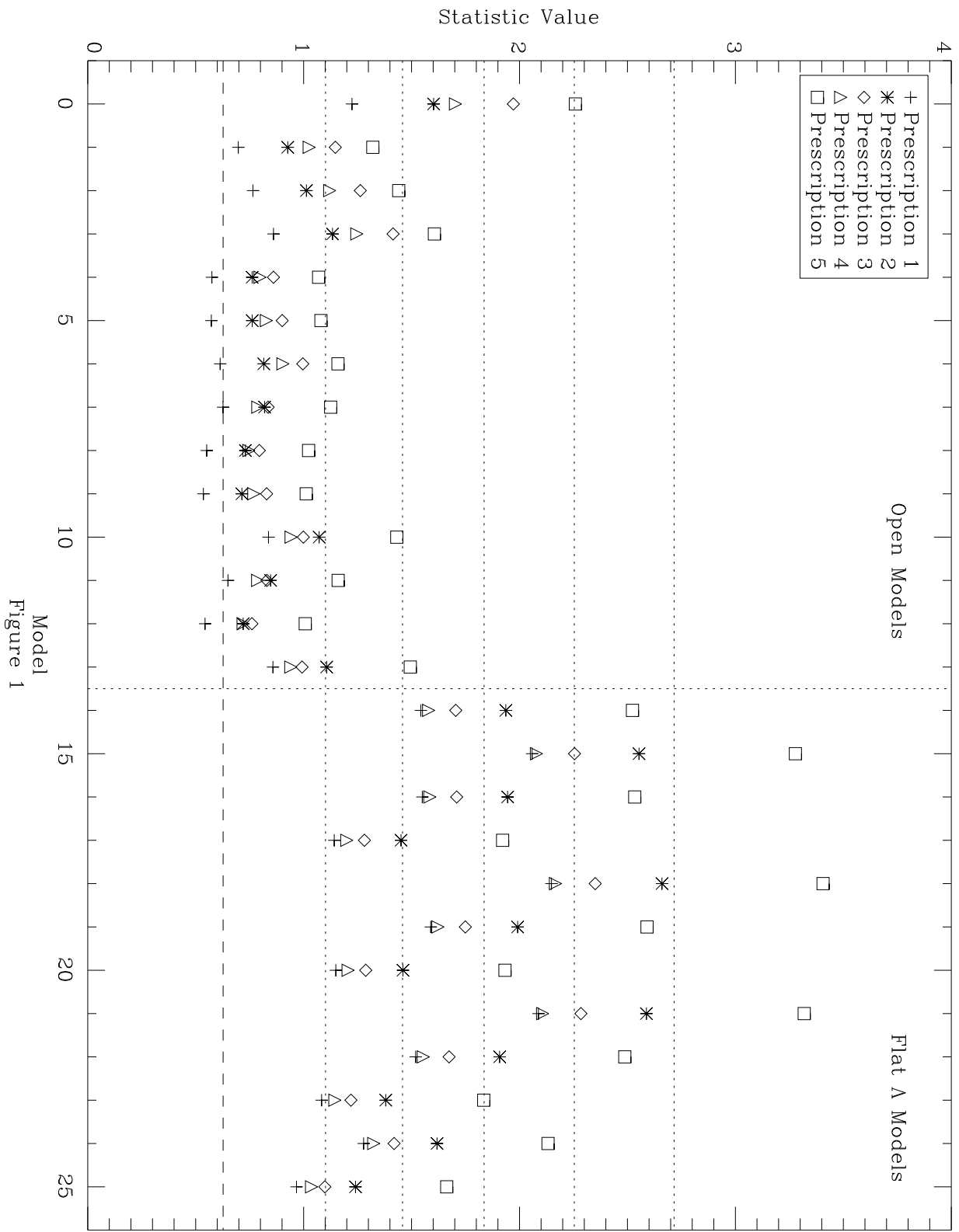

Figure 1

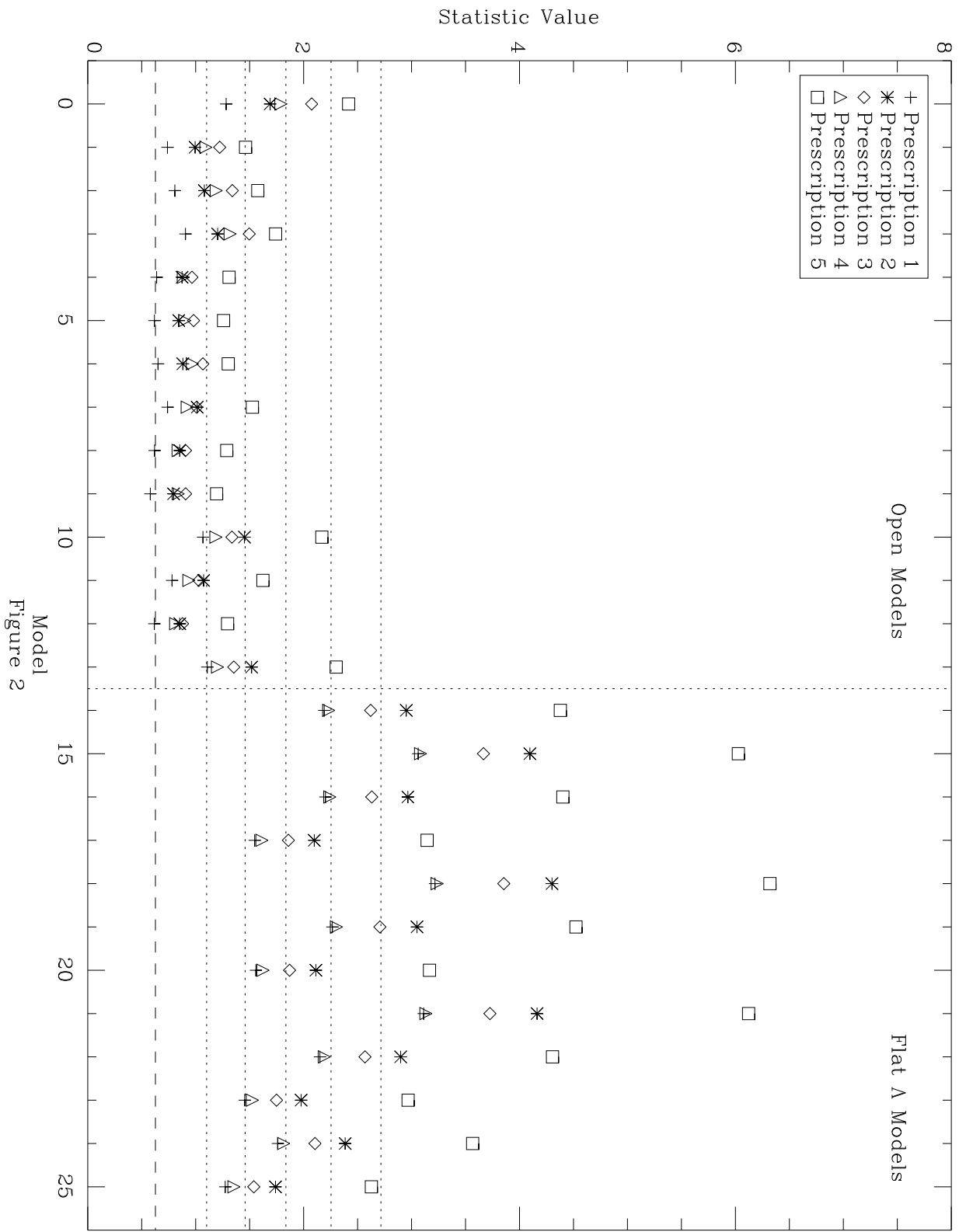

Figure 2

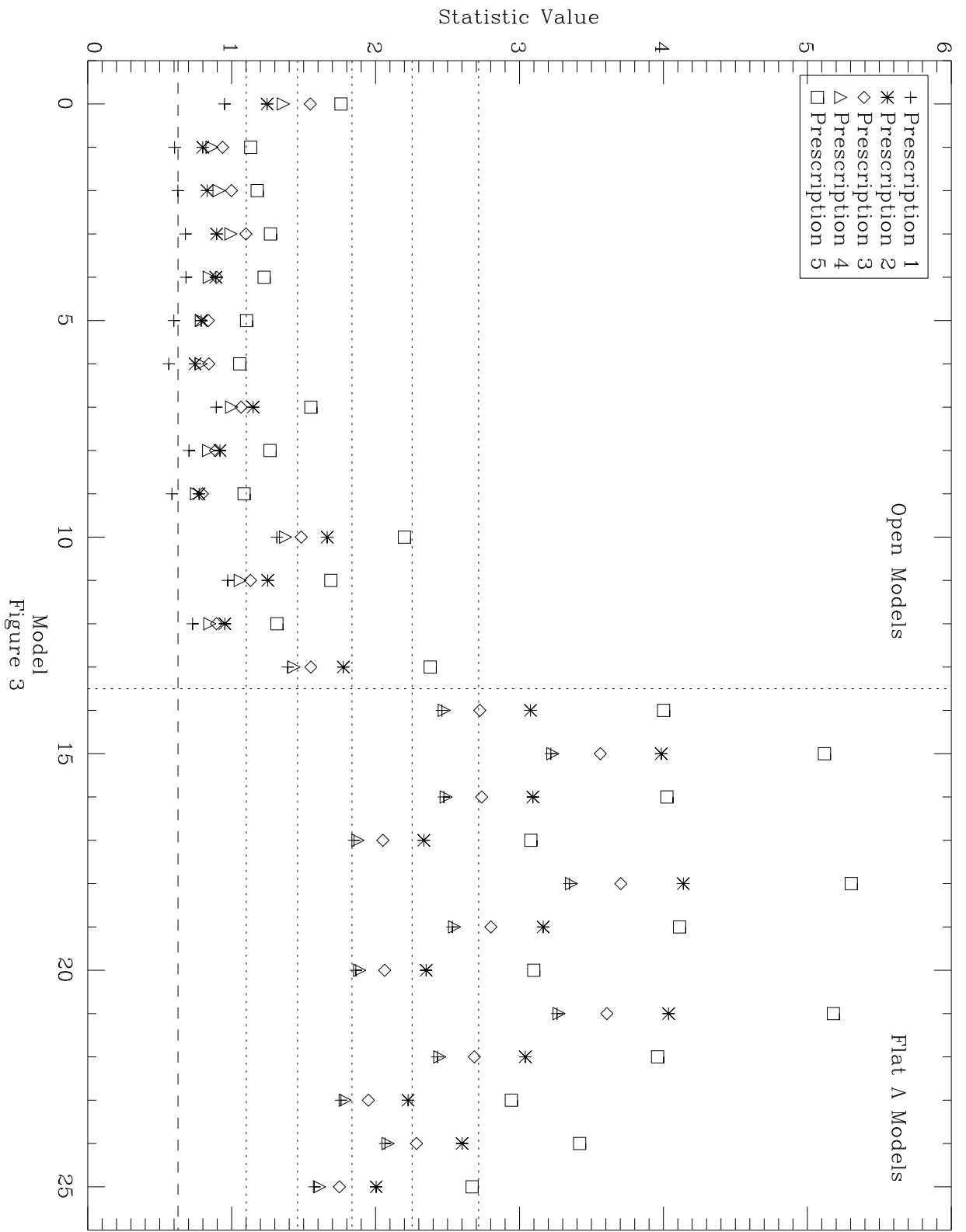

Figure 3

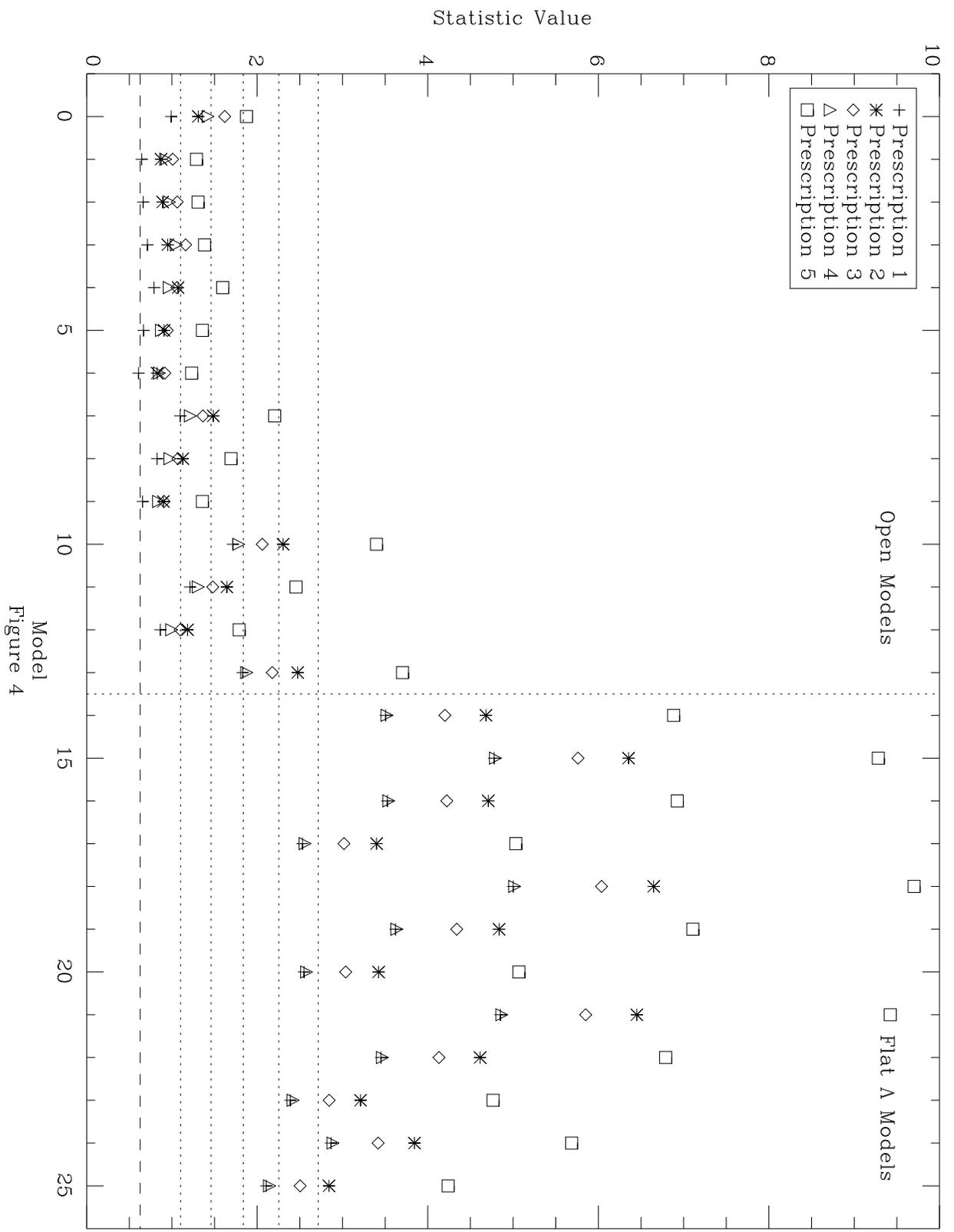

Figure 4

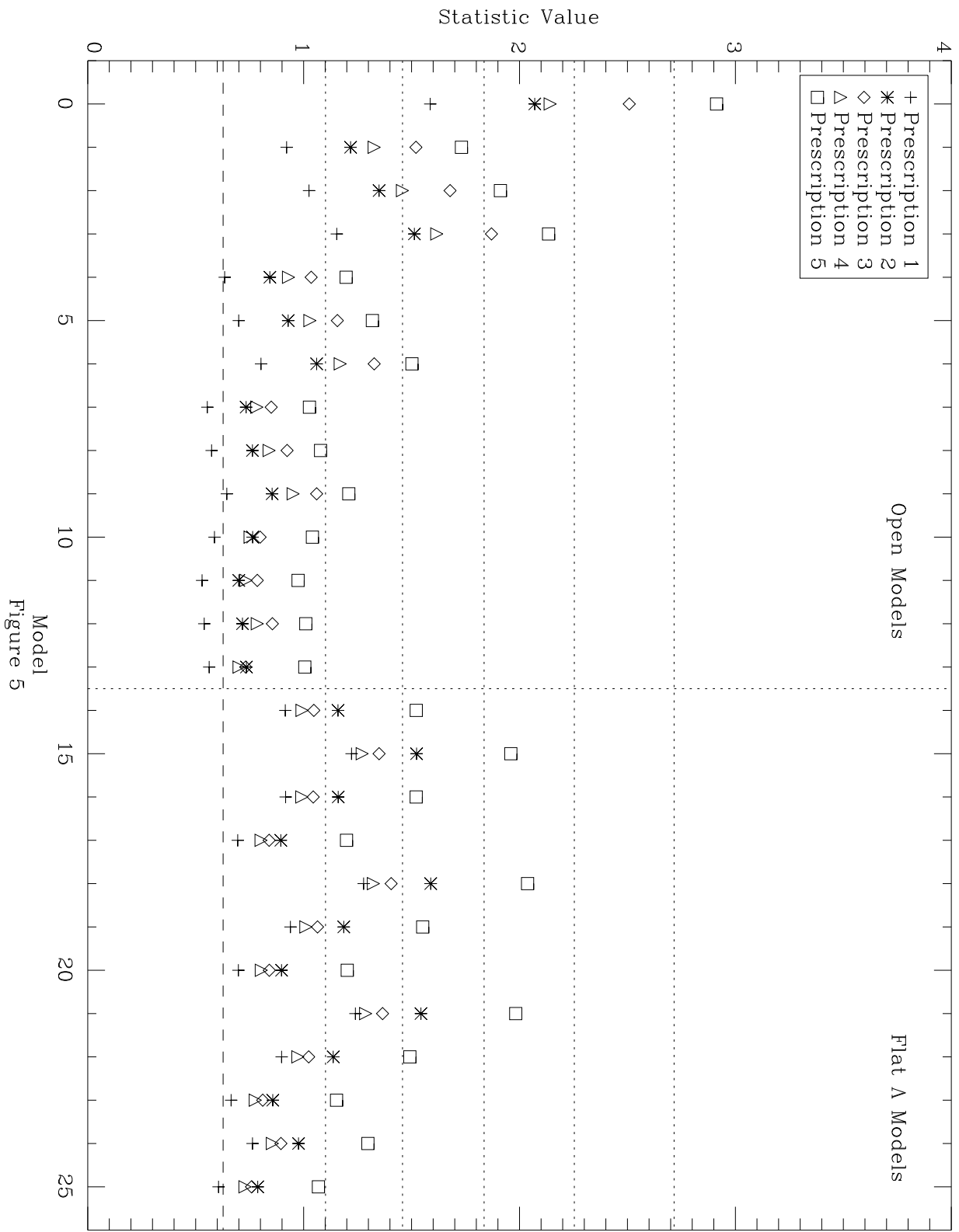

Figure 5

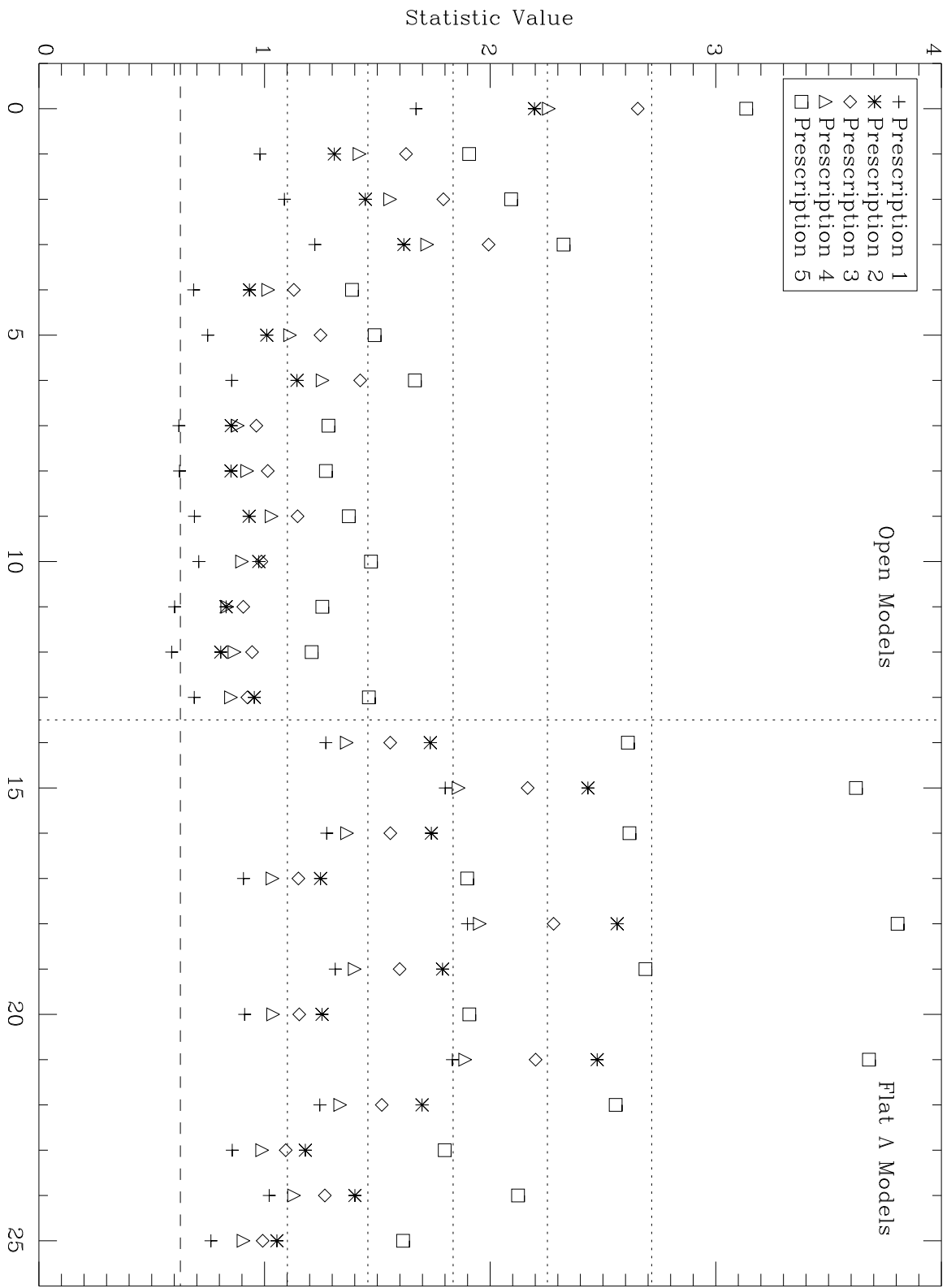

Figure 6